\documentclass[12pt]{amsart}
\usepackage{amsmath}
\usepackage{amssymb}
\advance \textheight by \topskip
\begin{document}
\bibliographystyle{abbrv}
\newtheorem{thm}{Theorem}[section]
\newtheorem{prop}[thm]{Proposition}
\newtheorem{defn}[thm]{Definition}
\newtheorem{lemma}[thm]{Lemma}
\newtheorem{cor}[thm]{Corollary}
\newtheorem{assume}[thm]{Assumption}
\newtheorem{example}[thm]{Example}
\newtheorem{theorem}{Theorem}
\newtheorem{corol}[theorem]{Corollary}
\newcommand{\real}{{\mathbb R}}
\newcommand{\com}{{\mathbb C}}
\newcommand{\quat}{{\mathbb H}}
\newcommand{\cayn}{{\mathbb O}}
\newcommand{\fld}{{\mathbb F}}
\newcommand{\reg}{G_{\mathbb R}}
\newcommand{\cog}{G_{\mathbb C}}
\newcommand{\sech}{{\mathrm{sech}}}
\newcommand{\prf}{\noindent{\bf Proof:\,\,}}
\newcommand{\myqed}{{\bf QED}}
\newcommand{\lieg}{{\frak g}}
\newcommand{\lieh}{{\frak h}}
\newcommand{\liep}{{\frak p}}
\newcommand{\liek}{{\frak k}}
\newcommand{\liegc}{\lieg_{\com}}
\newcommand{\p}{{\bf p}}
\newcommand{\roots}{\Phi}
\newcommand{\rhyp}{H^{n}_{\real}}
\newcommand{\chyp}{H^{n}_{\com}}
\newcommand{\qhyp}{H^{n}_{\quat}}
\newcommand{\cahyp}{H^{2}_{\cayn}}
\newcommand{\fhyp}{H^{n}_{\fld}}
\newcommand{\f}{_{\cal F}}
\newcommand{\ff}{_{\tilde{\cal F}}}
\newcommand{\fff}{\tilde{\cal F}}
\newcommand{\fg}{{\cal F}}
\newcommand{\m}{{\cal M}}
\newcommand{\mbar}{\overline{\cal M}}
\newcommand{\fbar}{\overline{\fg}}
\newcommand{\fh}{{\cal F}^{!}}
\newcommand{\ybar}{\overline{Y}}
\newcommand{\h}{{\cal H}}
\newcommand{\g}{{\cal G}}
\newcommand{\met}{{\cal C}(\eta_{1})}
\newcommand{\ab}{\mid}
\newcommand{\unab}{\mid}
\newcommand{\tbar}{\overline{T}}
\title[Morse theory and infinite families of harmonic maps]{}
\begin{center}
{\large\bf Morse theory and infinite families of harmonic maps\\
between spheres}\\[.1in]
{Kevin Corlette\\
Department of Mathematics\\
University of Chicago\\
5734 South University Avenue\\
Chicago, Illinois 60637}\\
{\tt kevin@math.uchicago.edu}\\[.1in]
{Robert M. Wald\\
Enrico Fermi Institute and Department of Physics\\
University of Chicago\\
5640 South Ellis Avenue\\
Chicago, Illinois 60637\\}
{\tt rmwa@midway.uchicago.edu}
\end{center}
\thanks{\hspace*{-.23in} The
first author was supported by NSF grant DMS-9971727 during
the course of this work. The second author was supported by NSF
grant PHY 95-14726.}
\begin{abstract}
Existence of an infinite sequence of harmonic maps between spheres of
certain dimensions was proven by Bizo\'{n} and Chmaj. This sequence shares
many features of the Bartnik-McKinnon sequence of solutions to the
Einstein-Yang-Mills equations as well as sequences of solutions that have
arisen in other physical models. We apply Morse theory methods to prove
existence of the harmonic map sequence
and to prove certain index and
convergence properties of this sequence. In addition, we generalize
the result of Bizo\'{n} and Chmaj to produce infinite sequences of
harmonic maps not previously known.  The key features ``responsible''
for the existence and properties of the sequence are thereby seen to be
the presence of a reflection (${\mathbb Z}_2$) symmetry and the existence of a
singular harmonic map of infinite index which is invariant under this symmetry.
\end{abstract}
\maketitle
\noindent
{\large\bf \S\thesection\,\,Introduction}\\[.1in]
\hspace*{.19in}
A countably infinite sequence, $\{ h_i \}$, of harmonic maps from an
$(m+1)$-dimensional sphere, $S^{m+1}$, into itself for $2 \leq m \leq 5$ was
discovered and analyzed by Bizo\'{n} \cite{MR97d:58056} and Bizo\'{n} and Chmaj
\cite{MR98d:58039}. In these references,
existence of this sequence was proven via a shooting
argument, and analytic arguments and/or numerical evidence also was
presented that this sequence satisfies the following properties: (1) As $i
\rightarrow \infty$, we have $h_i \rightarrow h_\infty$ pointwise
except at the poles, where $h_\infty$ denotes a singular harmonic map
which maps all of $S^{m+1}$ into the equator of $S^{m+1}$. (2) The sequence of
energies, $\{ E_i \}$, is monotone increasing and converges to the
energy, $E_\infty$, of $h_\infty$. (3) The index of $h_i$ is $i$.

The above properties bear a remarkable resemblance to the properties
of the Bartnik-McKinnon sequence and the colored black hole sequences
of Einstein-Yang-Mills theory (see Volkov and Gal'tsov \cite{volgal} for a
review). Here one considers static, spherically symmetric,
asymptotically flat, nonsingular solutions of the Einstein-Yang-Mills
equations where the static Killing field either remains strictly
timelike everywhere (the Bartnik-McKinnon case) or becomes null on a
regular event horizon (the colored black hole case). The
Bartnik-McKinnon solutions are labeled by a positive integer, $i$,
whereas the colored black holes are labeled by $i$ and a positive real
number, $r_0$, corresponding to the radius of the event horizon.  As
$i \rightarrow \infty$, the Bartnik-McKinnon sequence converges (in
the sense described in section 3.1 of \cite{volgal}) to the extreme
Reissner-Nordstrom solution with unit magnetic charge. At fixed $r_0$,
the colored black hole sequence converges (in the sense described
in section 4.1 of \cite{volgal}) to a unit magnetically charged
Reissner-Nordstrom solution. The mass of both the Bartnik-McKinnon and
colored black hole sequences increases monotonically with $i$ and
converges to the mass of the limiting Reissner-Nordstrom
solution. Finally, numerical evidence indicates that---if one suitably
restricts the function space so that only ``even parity''
perturbations (involving only variables that are nonzero in the
background) are considered---the index of both the $i$th Bartnik-McKinnon
and the $i$th colored black hole solution is $i$.
Sequences of solutions with similar properties have also been found in a
number of other models, in particular, static, spherically symmetric
solutions in Yang-Mills-dilaton theory \cite{b1} and self-similar wave
maps from Minkowski spacetime into $S^3$ (or, equivalently, harmonic maps
from the hyperboloid, $H^3,$ into $S^3$) \cite{b2}.

The fact that very similar sequences of solutions exist in the quite
different contexts of harmonic map theory and Einstein-Yang-Mills theory
suggests that there should be an explanation of the existence and
properties of these sequences of solutions that depends only on some
general properties of the equations, not on their detailed form. An early
attempt to provide such an explanation (made prior to the discovery of the
harmonic map sequence) was given in \cite{MR93k:83008}.
In that reference, it was
proposed that a key property related to the existence of the
Bartnik-McKinnon and colored black hole sequences is the presence of a
symmetry that ensures a ``degeneracy'' among solutions, i.e., that if a
solution of a given mass exists, then other solutions (obtained by action
of the symmetry on this solution) of the same mass also must exist. The
following heuristic argument for the existence of the Bartnik-McKinnon
sequence was given. (A similar argument also was given for the colored
black hole sequence.) Suppose that on the phase space, $\Gamma$,
appropriate to the problem there exists a ``mass flow'' vector field
(satisfying suitable smoothness properties and invariant under the
symmetries), such that the mass monotonically decreases along the integral
curves of this vector field, and such that these integral curves always
asymptotically approach a critical point of mass (corresponding to a
stationary solution of the Einstein-Yang-Mills equations
\cite{MR93k:83008}).
By
the positive mass theorem, the global minimum of mass is flat spacetime
with a pure gauge Yang-Mills field, as well as copies of this solution
under the symmetry. However, in addition to these global minima of mass
(and, possibly, other local minima of mass), there should exist other
critical points of mass: the integral curves of the mass flow vector field
should not be able to bifurcate discontinuously between the different
local minima, so there should exist points of phase space that do not flow
to any local minimum. Consequently, these must flow to critical points of
nonzero index. Indeed, if the set of points, $\Gamma_1$, that do not flow
to a local minimum of mass has the structure of a hypersurface of
co-dimension $1$, then the critical point of minimum mass within
$\Gamma_1$ would have index $1$, corresponding to the first
Bartnik-McKinnon solution. The argument can now be repeated, replacing
$\Gamma$ by $\Gamma_1$, to argue for existence of a critical point of
index $1$ within $\Gamma_1$ and, hence, index $2$ within the original
phase space $\Gamma$. It was proposed in \cite{MR93k:83008}
that this would account
for the second Bartnik-McKinnon solution; continued iteration of this
argument should generate the entire Bartnik-McKinnon sequence. This
argument would naturally account for the fact that the index of the $i$th
Bartnik-McKinnon solution is $i$, as well as for the fact that the masses
in the sequence increase monotonically.

However, in addition to its heuristic nature and its major gaps
concerning the existence of a suitable mass flow vector field, the
manifold nature of $\Gamma_i$, etc., the argument given in \cite{MR93k:83008}
suffers from the following three serious deficiencies: First, the
relevant symmetry was identified in \cite{MR93k:83008} as being the ``large
gauge transformations'' of the Yang-Mills field. However, no analog of
this symmetry exists in the harmonic map problem, so if a common
explanation is sought, this could not be the relevant symmetry in the
Einstein-Yang-Mills context. Second, no argument was given as to why
$\Gamma_1$ (or any of the higher $\Gamma_i$) should be connected
or---if not connected---why the symmetry should map any connected
component of $\Gamma_1$ into itself. If the symmetry fails to map a
connected component of $\Gamma_1$ into itself, no bifurcation of the
mass flow on $\Gamma_1$ need occur, and the above argument for
additional critical points breaks down. Note that this difficulty
would leave intact the argument for a critical point of index $1$
(i.e., the first Bartnik-McKinnon solution) since $\Gamma$ is
connected, but if the relevant symmetry is taken to be the large
gauge transformations, there is no reason to expect that any higher
members of the Bartnik-McKinnon sequence need exist. Third, the argument
given in \cite{MR93k:83008} does not account for any of the convergence
properties of the Bartnik-McKinnon sequence as $i \rightarrow \infty$.

In this paper, we will give a Morse theoretic proof of the existence and
properties of a generalization of the sequence of
harmonic maps between spheres found in
\cite{MR97d:58056} and \cite{MR98d:58039}.
By doing so, we will---in the context of the harmonic
map problem rather than the Einstein-Yang-Mills problem---in effect, cure
all of the deficiencies as well as close all of the gaps in the general
argument sketched in \cite{MR93k:83008}. In our proofs,
the relevant symmetry on
the space of maps between spheres will be seen to be a ${\mathbb Z}_{2}$
symmetry
corresponding to composing a given map between the spheres with the
reflection isometry about the equatorial plane of the image sphere. It
will be seen that an additional fact playing a crucial role in our proofs
is the existence of a harmonic map (namely, $h_\infty$) which is invariant
under this symmetry and which has infinite index. At a heuristic level,
the presence of $h_\infty$ ensures that the heat flow has the appropriate
bifurcation properties to obtain an infinite sequence of solutions.
However, a number of technical difficulties would arise if we attempted to
use heat flow arguments in our proofs. The heat flow for maps into a sphere
can develop singularities in finite time, which makes it difficult
to regard it as a flow on the space of maps.  As is typical in
Morse theoretic arguments on infinite-dimensional spaces, we replace
the heat flow by an energy-decreasing
flow which is defined in a more {\it ad hoc}
manner, but has the two advantages that it is defined for all time
and its flow lines converge to limiting maps as time tends to infinity.
It then becomes possible to apply the essential observation of
Morse theory, which is that nontrivial topology in the space of
maps forces the existence of critical points for the energy
which are not minima. More precisely, one can apply a minimax
argument for each homology class in the configuration space: the minimum
of the collection of numbers $E$ such that
the chosen homology class can be realized in the space of maps
with energy no larger than $E$ is a critical value.  However,
the space of maps in our case is parametrized
by real-valued functions on the real line, so has no nontrivial
topology.  We avoid this problem by (1) showing that all critical
points have energy bounded by the energy of the unique singular
critical point (corresponding to a map which collapses the
domain sphere to the equator in the target sphere) and (2)
exploiting the ${\mathbb Z}_{2}$ symmetry.  This leads to
a configuration space with nontrivial homology
classes in infinitely many dimensions.  These homology
classes are the essential explanation for the existence of
the infinite sequence of harmonic maps. 

Several technical complications arise in the course of the
proof which may obscure the main ideas, so we will outline
a naive version of the argument here.  The first step is
to reduce the harmonic map equation to an ordinary differential
equation by prescribing the map along codimension one slices
in the domain sphere.  This leads to the condition that the
map along such slices must be an eigenmap, i.e.\ a harmonic
map with constant energy density.  By an appropriate choice
of coordinates, we can then reduce the problem to that of
finding critical points of an energy functional $E$ on a
Hilbert space $H$
of functions $h:{\mathbb R}\rightarrow{\mathbb R}.$  There is
an involution of $H$ given by multiplication by $-1$ which
preserves $E.$  It has a unique fixed point, corresponding
to the function $h_{\infty}$ which is identically zero, and
corresponding to a singular harmonic map which collapses
the domain to the equator of the target.  We can show that
all other critical points of $E$ correspond to smooth
harmonic maps with energy strictly less than $E(h_{\infty}).$
Under suitable conditions, $h_{\infty}$ has infinite index
as a critical point of $E.$  Ideally, we would exploit this
as follows.  Consider the punctured Hilbert space $H^{*}=
H-\{h_{\infty}\},$ and divide by the involution to obtain
a space $\overline{H^{*}}$ with the homotopy type of
${\mathbb R}P^{\infty}.$  In particular, the homology of
$\overline{H^{*}}$ is ${\mathbb Z}_{2}[x],$ where $x$ is a
class of degree 1.  In the cases where $h_{\infty}$ has
infinite index, we can realize all of the homology of
$\overline{H^{*}}$ in the portion consisting of functions
with energy less than $E(h_{\infty}).$  If Morse theory
could be applied naively, it would tell us that there must
be an infinite sequence of critical points, with at least
one critical point of index $k$ for every $k\geq0.$
The idea of finding harmonic maps between
spheres by reducing to an ordinary differential equation is
an old one, first used by Smith \cite{MR45:7642}\cite{MR52:11949}
and used more
recently by Ding \cite{MR90b:58049}, Eells-Ratto \cite{MR94k:58033} and
Pettinati-Ratto \cite{MR91g:58063}.  To the best of our knowledge, most of
the work done in this direction has focused on showing that harmonic
maps exist, without trying to show that they exist in profusion.
The work of Bizo\'{n} and Bizo\'{n}-Chmaj appears to be the first
in this direction.

It should be noted that the static, spherically symmetric
Einstein-Yang-Mills equations also possess a ${\mathbb Z}_2$ symmetry given by
the transformation $w \rightarrow -w$ in the notation of \cite{volgal}
(see eqs.(2.50)-(2.53) of that reference). In addition, the
Reissner-Nordstrom solutions with unit magnetic charge are invariant
under this symmetry and have infinite index. This suggests that it
might be possible to give a similar Morse theoretic proof of the
existence and properties of the Bartnik-McKinnon and colored black
hole sequences; the main problem would be to show that the
relevant functional satisfies the Palais-Smale condition.
However, we shall not pursue this issue here.

The contents of this paper are as follows.  In \S1, we describe
the basic framework for studying special classes of harmonic maps
which can be interpreted as solutions of ordinary differential
equations.  We establish certain basic properties of these solutions,
including the bound on the energy mentioned above.  In \S2, we study
the index of the singular map which collapses the domain sphere
onto the equator of the target, and show that it is infinite
under certain circumstances.  In the final section, we apply
Morse theory for functions on compact
convex sets to establish the existence of an infinite sequence of
harmonic maps.

Both authors would like to thank Piotr Bizo\'{n} for helpful
discussions related to this paper.  The second author would
particularly like to acknowledge conversations with Bizo\'{n}
several years ago which provided the initial impetus for this project.
\newpage
\addtocounter{section}{1}
\noindent
{\large\bf \S\thesection\,\,Preliminaries}\\[.1in]
\hspace*{.19in}
Let $S^{n}$ be the n-dimensional sphere with the Riemannian metric
induced from its identification with the unit sphere in Euclidean
space.  We identify the $(n+1)$-sphere with north and south poles
removed with
\begin{equation}(-\frac{\pi}{2},\frac{\pi}{2})\times S^{n}\end{equation}
with the metric
\begin{equation}d\theta^{2}+\cos^{2}{\theta}\,d\psi^{2},\end{equation}
where $\theta$ is the coordinate along
$(-\frac{\pi}{2},\frac{\pi}{2})$
and $d\psi^{2}$ is the round metric on $S^{n}.$
If we fix a map $F:S^{m}\rightarrow S^{n},$
then a function
$f:[-\frac{\pi}{2},\frac{\pi}{2}]\rightarrow[-\frac{\pi}{2},\frac{\pi}{2}]$
with $f(\pm\frac{\pi}{2})=\pm\frac{\pi}{2}$ gives a map
$\tilde{f}:S^{m+1}\rightarrow S^{n+1}$ via
$\tilde{f}(\theta,\psi)=(f(\theta),F(\psi)).$

We are interested in harmonic maps which can be written in this
form with $m\geq2.$
In order for this to be reasonable, we must impose a condition
on $F.$
\begin{defn}
$F:S^{m}\rightarrow S^{n}$ is an eigenmap if it is harmonic and the
energy density $|dF|^{2}=\omega$ is constant as a function on $S^{m}.$
In this situation, $\omega$ is called the eigenvalue of the map.
\end{defn}
One of the simplest examples of an eigenmap is the identity map
$S^{m}\rightarrow S^{m};$ the corresponding eigenvalue is $m.$
Other examples are given by the Hopf maps $S^{3}\rightarrow S^{2},$
$S^{7}\rightarrow S^{4}$ and $S^{15}\rightarrow S^{8};$ these
have eigenvalues 8, 16 and 32, respectively.  Any eigenmap
$S^{m}\rightarrow S^{n}$ is obtained from a collection of
$n+1$ eigenfunctions $\xi_{1},\ldots,\xi_{n+1}$ of the Laplacian
on $S^{m}$ satisfying $\sum\xi_{i}^{2}=1.$  There is no general
classification of eigenmaps between spheres, but a number of
examples and partial results are known.  For example, there
are eigenmaps produced by what is known as the
Hopf construction, generalizing the classical Hopf maps.
There are also eigenmaps associated with harmonic
eiconals.  These are harmonic polynomials on ${\mathbb R}^{n+1}$
whose gradients have unit length along the unit sphere;
the gradients then give harmonic maps $S^{n}\rightarrow
S^{n}.$  Consult Eells-Ratto
\cite{MR94k:58033}, Chapter VIII for more information.

Once we assume that $F$ is an eigenmap, the condition for $\tilde{f}$
to be a harmonic map
reduces to a differential equation for $f:$
\begin{equation}f''-m\tan{\theta}f'+\frac{\omega}{2}\sec^{2}{\theta}\sin{2f}=0.\end{equation}
This is the Euler-Lagrange equation for the following energy
functional:
\begin{equation}J(f)=\frac{1}{2}\int_{-\frac{\pi}{2}}^{\frac{\pi}{2}}[(f')^{2}+
\frac{\omega\cos^{2}{f}}{\cos^{2}{\theta}}]\cos^{m}{\theta}\,d\theta,\end{equation}
which gives the energy of $\tilde{f}$ in the usual sense.
It will often be useful to make the change of variables
$x=\log{(\tan{\frac{1}{2}(\theta+\frac{\pi}{2})})}$ and to set
$h(x)=f(2\tan^{-1}(e^{x})-\frac{\pi}{2}).$  Then the energy
becomes
\begin{equation}E(h)=\frac{1}{2}\int_{-\infty}^{\infty}[(h')^{2}+\omega\cos^{2}{h}]
\sech^{m-1}x\,dx,\end{equation}
while the Euler-Lagrange equation becomes
\begin{equation}h''-(m-1)\tanh{x}h'+\frac{\omega}{2}\sin{2h}=0.
\label{eq:eulag}\end{equation}
We will denote the map of spheres associated to $h$ by $\tilde{h}.$
If $v:\real\rightarrow\real$ is $C^{2}$ with compact support, then
the formula for the second variation of $E$ is
\begin{equation}\begin{array}{rcl}
\frac{d^{2}\,}{dt^{2}}E(h+tv)|_{t=0} & = & \int_{-\infty}^{\infty}
[(v')^{2}-\omega\cos{2h}v^{2}]\sech^{m-1}x\,dx\\
& = & -\int_{-\infty}^{\infty}[(v'\sech^{m-1}x)'+\omega\cos{2h}v
\sech^{m-1}x]v\,dx\\
& = & -\int_{-\infty}^{\infty}[v''-(m-1)\tanh{x}v'+\omega\cos{2h}v]v
\sech^{m-1}x\,dx.
\end{array}\end{equation}
The first of these integrals will be abbreviated as $Q(v,v).$

Define a weighted Sobolev space $H$ to be the completion
of the space of smooth functions
$h:\real\rightarrow\real$ satisfying
\begin{equation}
\int_{-\infty}^{\infty}[(h')^{2}+h^{2}]\sech^{m-1}x\,dx<\infty\end{equation}
with respect to the norm $\parallel h\parallel_{H}^{2}$
defined by the integral above.
Define $E:H\rightarrow\real$ by the same formula as in the previous
paragraph.  $E$ is
easily seen to be a smooth function on $H,$ with critical points
given by functions satisfying the Euler-Lagrange equation given
above.  We are interested in critical points of $E$ lying
in the closed convex set $C\subset H$ given by
\begin{equation}C=\{h\in H||h(x)|\leq\frac{\pi}{2},\,x\in\real\}.
\end{equation}  The fact that this set is closed follows from
the Sobolev embedding theorem for ordinary Sobolev spaces
on a finite interval.
We begin by proving a few simple properties of these critical
points.  Most, but not all, of this can be found in
\cite{MR97d:58056}\cite{MR98d:58039}.  In fact, we will need
to study critical points of functionals which are perturbations
of the energy, so define
\begin{equation}E_{\nu}(h)=\frac{1}{2}\int_{-\infty}^{\infty}[(h')^{2}+\omega
(1+\nu)\cos^{2}{h}]\sech^{m-1}{x}\,dx,\end{equation}
where $\nu$ is a $C^{2}$ function on $\real$ with
$|\nu|<1$ and
compact support.
The Euler-Lagrange equation for this functional is
\begin{equation}h''-(m-1)\tanh{x}h'+\frac{\omega}{2}(1+\nu)\sin{2h}=0.
\label{eq:eulag1}\end{equation}

Given a solution $h\in C$ of this equation, define
\begin{equation}W(x)=\frac{1}{2}(h')^{2}+\frac{\omega}{2}(1+\nu)\sin^{2}{h}.\end{equation}
We calculate that
\begin{equation}\begin{array}{rcl}
\frac{dW}{dx} & = & h''h'+\omega(1+\nu)\sin{h}\cos{h}h'+\frac{\omega}{2}
\nu'\sin^{2}{h}\\
 & =& (h''+\frac{\omega}{2}(1+\nu)\sin{2h})h'+\frac{\omega}{2}
\nu'\sin^{2}{h}\\
& = & (m-1)\tanh{x}(h')^{2}+\frac{\omega}{2}
\nu'\sin^{2}{h}.
\end{array}\end{equation}
Thus, $W$ is increasing when $x\gg 0,$ and decreasing when
$x\ll 0.$  If $W(x_{0})\geq\frac{\omega}{2}$ for some $x_{0}\gg 0,$
then either $h$ is constant with value $\pm\frac{\pi}{2}$ or
$h'(x)>\epsilon>0$ for all $x>x_{0}.$  The latter is impossible,
since $h\in C.$  A similar argument applies if $x_{0}\ll 0.$
Hence, $W(x)\leq\frac{\omega}{2}$ for all $x$ sufficiently
far from zero.
As a consequence, the limits
\begin{equation}\lim_{x\rightarrow\pm\infty}W(x)=L_{\pm}\end{equation}
both exist,
and
\begin{equation}\lim_{x\rightarrow\pm\infty}W'(x)=0.\end{equation}
The latter implies that $h'$ approaches zero as $x$ tends
to $\pm\infty.$  Taken together, the fact that both $W$ and
$h'$ have limits at infinity implies that
\begin{equation}\lim_{x\rightarrow\pm\infty}\sin^{2}{h}=\frac{2L_{\pm}}{\omega},\end{equation}
and therefore $h$ itself has limits at $\pm\infty.$  The Euler-Lagrange
equation now implies that $h''$ also has limits at $\pm\infty,$
and since $\lim_{x\rightarrow\pm\infty}h'(x)=0,$ that limit must
be zero.  The Euler-Lagrange equation then implies that
\begin{equation}\lim_{x\rightarrow\pm\infty}\sin{2h}=0,\end{equation} which means that
$h$ approaches $0$ or $\pm\frac{\pi}{2}$ as $x$ approaches $\pm\infty.$
If the limit is zero at either extreme, then $W$ approaches zero as
well.  Since it is
increasing when $x\gg 0$ and decreasing when
$x\ll 0,$ and nonnegative everywhere,
this implies that it must be zero for $x\gg 0$ or $x\ll 0.$
The latter implies that $h=0$ on an open set, so is
zero everywhere.  Thus,
unless $h=h_{\infty},$ 
\begin{equation}\lim_{x\rightarrow\pm\infty}h(x)=\pm\frac{\pi}{2}.\end{equation}
This implies that the corresponding map $\tilde{h}$
between spheres
is continuous at the poles of $S^{m+1}.$ The Euler-Lagrange
equation for $E_{\nu}$ corresponds to harmonic map equation
in a neighborhood of each pole, so by regularity of
continuous solutions of the harmonic map equation, $\tilde{h}$ must
be a smooth map.

It is also possible to compare the perturbed
energy of
$h$ with that of $h_{\infty}.$
When $m\leq1,$ $h_{\infty}$ has infinite energy, so
from here on we will assume $m\geq2.$
Integrating by parts, and using the fact that $h'$
tends to zero at $\pm\infty,$ we find that
\begin{equation}\begin{array}{rcl}
E_{\nu}(h) & = & \frac{1}{2}\int_{-\infty}^{\infty}h[-h'\sech^{m-1}x]'
+\omega(1+\nu)\cos^{2}{h}\sech^{m-1}{x}dx\\
& = & \frac{1}{2}\int_{-\infty}^{\infty}[\frac{1}{2}h\sin{2h}
+\cos^{2}{h}]\omega(1+\nu)\sech^{m-1}{x}dx.\end{array}\end{equation}
The function $\frac{1}{2}h\sin{2h}+\cos^{2}{h}$ is bounded by
$1$ when $h\in C,$ and is identically equal to $1$ only if $h=h_{\infty}.$
Thus, $E_{\nu}(h)\leq E_{\nu}(h_{\infty}),$ with equality only if
$h=h_{\infty}.$

We summarize this in the following
\begin{prop}
Assume that $m\geq2.$
Any critical point $h\in C$ of $E_{\nu}$
is either the singular map $h_{\infty}$
or satisfies
\begin{enumerate}
\item $\lim_{x\rightarrow\infty}h(x)=\pm\frac{\pi}{2},\,
\lim_{x\rightarrow -\infty}h(x)=\pm\frac{\pi}{2}$
\item $E_{\nu}(h)<E_{\nu}(h_{\infty})$
\item The index and nullity of the Hessian of $E_{\nu}$ at $h$
are finite.
\end{enumerate}
\end{prop}
\prf The only remaining issue is the proof of (3).  The critical
points of $E_{\nu}$ correspond to harmonic maps with a potential
function from $S^{m+1}$ to $S^{n+1}.$  As such, they satisfy
a quasilinear system of elliptic PDE which, in a neighborhood
of each pole, agrees with the usual
harmonic map equation.
One can calculate the Hessian of the perturbed energy functional
on the space of all maps $S^{m+1}\rightarrow S^{n+1};$ it
corresponds to an elliptic operator which, aside from a zeroth
order term deriving from the potential, agrees with the
corresponding operator for the usual energy.  This is an
operator of Laplacian type, so the assertions about the
index and nullity of the Hessian follow easily. \myqed 

It is of interest to consider functions satisfying certain
symmetry conditions:  either
\begin{equation}
h(-x)=h(x)\label{eq:degree0}\end{equation}
or
\begin{equation}
h(-x)=-h(x).\label{eq:suspension}\end{equation}
If $h$ satisfies condition 1.2.1 in the
previous Proposition, then \ref{eq:degree0} implies
that the corresponding map between spheres is homotopically
trivial, while \ref{eq:suspension} implies that the map between
spheres is in the homotopy class of the suspension of
$F.$  We will let $H^{+}$ be the set of functions in
$H$ satisfying (\ref{eq:degree0}), and $H^{-}$ the set of functions in
$H$ satisfying (\ref{eq:suspension}); similarly,
$C^{+}=C\cap H^{+}$ and $C^{-}=C\cap H^{-}.$  Notice that, if
$\nu(-x)=\nu(x)$ (as we shall assume henceforth),
then the lefthand side of (\ref{eq:eulag1})
satisfies (\ref{eq:degree0}) if $h$ does, and it satisfies
(\ref{eq:suspension}) if $h$ does.  This implies that any
critical point of $E_{\nu},$ regarded as a function on
$H^{+}$ or $H^{-},$ is actually a critical point of $E_{\nu}$
regarded as a function on $H.$
\addtocounter{section}{1}\setcounter{thm}{0}
\\[.17in]
{\large\bf \S\thesection\,\,The index of the singular map}\\[.1in]
\hspace*{.19in}
In this section, we will show that the index of the singular map
corresponding to $h_{\infty}$ is infinite in certain cases.
The Hessian of $E_{\nu}$ at $h$ is given by \begin{equation}Q_{\nu}(v,v)=
\int_{-\infty}^{\infty}[(v')^{2}-\omega(1+\nu)\cos{(2h)}v^{2}]\sech^{m-1}xdx\end{equation}
for any $v\in H.$  Let 
\begin{equation}w=v\sech^{\frac{m-1}{2}}x.\end{equation}  Notice that if $v\in H,$
then
\begin{eqnarray}
\lefteqn{\int_{-\infty}^{\infty}[(w')^{2}+w^{2}]dx}\nonumber \\ 
 &= & \int_{-\infty}^{\infty}
[(v')^{2}+v^{2}]\sech^{m-1}{x}dx\\ & &\mbox{} +\int_{-\infty}^{\infty}[\frac{(m-1)^{2}}{4}
\sech^{m-1}{x}\tanh^{2}{x}v^{2}-(m-1)\sech^{m-1}{x}\tanh{x}vv']dx\nonumber
\\
&=  & \int_{-\infty}^{\infty}
[(v')^{2}+v^{2}]\sech^{m-1}{x}dx-\int_{-\infty}^{\infty}[
\frac{m-1}{2}\sech^{m-1}{x}\tanh{x}v^{2}]'dx\nonumber \\ & &
\mbox{}+\int_{-\infty}^{\infty}[
\frac{m-1}{2}[\sech^{m-1}{x}\tanh{x}]'v^{2}+\frac{(m-1)^{2}}{4}
\sech^{m-1}{x}\tanh^{2}{x}v^{2}]dx.\nonumber\end{eqnarray}
The middle term in the last expression vanishes, while the third is comparable
to
\begin{equation}\int_{-\infty}^{\infty}v^{2}\sech^{m-1}{x}dx,\end{equation}
so $v\in H$ if and only if $w\in L^{2}_{1},$ the usual
Sobolev space of $L^{2}$ functions on $\real$ whose first
derivatives are also in $L^{2}.$
A calculation shows that the 
integral defining the Hessian becomes
\begin{equation}\int_{-\infty}^{\infty}[(w'+\frac{m-1}{2}\tanh{x}w)^{2}-
\omega(1+\nu)\cos{(2h)}w^{2}]dx\end{equation}\begin{displaymath}
=\int_{-\infty}^{\infty}[(w')^{2}+
(m-1)\tanh{x}ww'+(\frac{(m-1)^{2}}{4}\tanh^{2}{x}-
\omega(1+\nu)\cos{(2h)})w^{2}]dx\end{displaymath}
\begin{eqnarray*}\lefteqn{=\int_{-\infty}^{\infty}[(w')^{2}+\frac{m-1}{2}(w^{2}\tanh{x})'
-\frac{m-1}{2}w^{2}\sech^{2}{x}}\\& &\mbox{}+(\frac{(m-1)^{2}}{4}\tanh^{2}{x}-
\omega(1+\nu)\cos{(2h)})w^{2}]dx\end{eqnarray*}
\begin{displaymath}=\int_{-\infty}^{\infty}[(w')^{2}+[\frac{(m-1)^{2}}{4}-
(\frac{(m-1)^{2}}{4}+\frac{m-1}{2})\sech^{2}{x}-
\omega(1+\nu)\cos{(2h)}]w^{2}]dx.\end{displaymath}
Now if we set $h=h_{\infty}=0,$ 
we find that the Hessian at $h_{\infty}$ is given by
\begin{equation}{\mathcal{Q}}_{\nu}(w,w)=\int_{-\infty}^{\infty}[(w')^{2}+[\frac{(m-1)^{2}}{4}-
(\frac{(m-1)^{2}}{4}+\frac{m-1}{2})\sech^{2}{x}-
\omega(1+\nu)]w^{2}]dx.\end{equation}
\begin{thm}
If $\frac{(m-1)^{2}}{4}<\omega,$ then there are finite-dimensional
subspaces of $H$ of arbitrarily large dimension on which
the Hessian of $E_{\nu}$
at $h_{\infty}$ is negative definite.
\end{thm}
\prf Under the assumption that $\frac{(m-1)^{2}}{4}<\omega,$
we
can find an $\epsilon>0$ and a $K>0$ such that
\begin{equation}V(x)=[\frac{(m-1)^{2}}{4}-
(\frac{(m-1)^{2}}{4}+\frac{m-1}{2})\sech^{2}{x}-
\omega(1+\nu)]<-\epsilon\end{equation}
whenever $|x| > K.$

Choose $a,c>0$ and consider the piecewise linear test function
\begin{equation}
F(x)  = \left\{\begin{array}{l}  0,\,\, x\not\in[c,c+2a]\\
   x-c,\,\,x\in[c,c+a]\\
 c+2a-x,\,\,x\in(c+a,c+2a].
\end{array}\right.\end{equation}
Then $F$ is in the Sobolev space $L^{2}_{1},$
and it is easy to calculate that
\begin{equation}{\mathcal{Q}}_{\nu}(F,F)=2a+\int_{-\infty}^{\infty}V(x)F(x)^{2}\,dx.\end{equation}
Choosing $c>K$ (so that $V(x)<-\epsilon$ throughout the support
of $F),$ we find that
\begin{equation}{\mathcal{Q}}_{\nu}(F,F)<2a-\epsilon\int_{-\infty}^{\infty}F(x)^{2}\,dx=
2a-\frac{2\epsilon a^{3}}{3}.\end{equation}
Thus, if $a^{2}>3\epsilon^{-1},$ then ${\mathcal Q}_{\nu}(F,F)<0.$
Given some $a$ satisfying this condition, for any positive integer
$i,$ define $F_{i}$ as above with $c=K+2ai.$  Then ${\mathcal Q}_{\nu}$
is negative definite on any subspace of $L^{2}_{1}$ generated by
any finite collection of the $F_{i}.$  \myqed

The identity map satisfies the hypothesis of this result for
$2\leq m\leq 5.$  The Hopf maps for which it holds are
$S^{3}\rightarrow S^{2}$ and $S^{7}\rightarrow S^{4}.$
Among other maps produced by the Hopf construction, there
are maps $S^{5}\rightarrow S^{4}$ and $S^{9}\rightarrow S^{8}$
which satisfy the hypothesis.  There are also maps associated
to harmonic eiconals to which the result applies.  For example,
there are harmonic eiconals of polynomial degree 3 on ${\mathbb R}^{5},$
${\mathbb R}^{8},$ ${\mathbb R}^{14}$ and ${\mathbb R}^{26},$
corresponding to harmonic self-maps of $S^{4},$ $S^{7},$ $S^{13}$
and $S^{25}$ with Brouwer degrees 0, 2, 2 and 2 and
eigenvalues 18, 27, 45 and 81, respectively.  The first three
satisfy the hypothesis.

Notice that $h_{\infty}\in H^{+},H^{-}.$  It can be shown that
the index of $h_{\infty}$ as a critical point of $E_{\nu}$ on
either of these spaces continues to be infinite under the same
hypothesis on $m$ and $\omega.$  This follows by a small modification
of the argument above, where the piecewise linear function $F$ given
there is replaced by $F(x)+F(-x)$ in the case of $H^{+},$ and
$F(x)-F(-x)$ in the case of $H^{-}.$ 
\addtocounter{section}{1}\setcounter{thm}{0}
\\[.17in]
{\large\bf \S\thesection\,\,Morse theory applied}\\[.1in]
\hspace*{.19in}
It is now possible to apply the elements of Morse theory on
convex sets to the $E_{\nu}.$  As basic references on this subject,
we take Chang \cite{MR94e:58023} and Struwe \cite{MR90h:58016}.
We first recall the notion of a critical point of a function on
a convex set (\cite{MR94e:58023}, Definition 6.4 or
\cite{MR90h:58016},
II,1.3).
\begin{defn}
$h_{0}\in C$ is a $C$-critical point of $E_{\nu}$ if
\begin{equation}dE_{\nu}(h_{0})(h-h_{0})\geq 0\end{equation}
whenever $h\in C.$  Equivalently, let
\begin{equation}g_{\nu}(h_{0})=\inf dE_{\nu}(h_{0})(h-h_{0}),\end{equation}
where $h$ ranges over elements of $C$ with $\parallel
h-h_{0}\parallel_{H}<1.$
Then $h_{0}$ is a C-critical point for $E_{\nu}$ iff $g_{\nu}(h_{0})=0.$
\end{defn}
It should be noted that $g_{\nu}$ is a continuous function on $C.$
We can define the notions of $C^{+}$-critical and $C^{-}$-critical
points of $E_{\nu}$ on $C^{+}=C\cap H^{+}$ and $C^{-}=C\cap H^{-};$
it is simply necessary to let $h$ range over elements of $C^{\pm}$
rather than elements of $C.$  It is not hard to see that any
$C^{\pm}$-critical point is also $C$-critical, since the differential
of $E_{\nu}$ at $h$ satisfies the same symmetry condition as $h.$ 

We now
compare the $C$-critical points with ordinary ones.
\begin{lemma}
Any $C$-critical point $h_{0}$ of $E_{\nu}$ is a critical point for
$E_{\nu}$ as a function on $H.$
\end{lemma}
\prf  If $h_{0}$
is not identically equal to $\pm\frac{\pi}{2},$ then there is a nonempty
open set of $\real$ on which it takes values in $(-\frac{\pi}{2},
\frac{\pi}{2}).$  On this open set, we can test $h_{0}$ by smooth
variations with compact support; when these are sufficiently small,
we do not leave $C.$  Hence, $h_{0}$ satisfies the Euler-Lagrange
equation for $E_{\nu}$ on the open set where it is not equal to
$\pm\frac{\pi}{2}.$

Near any point $x_{0}\in\real$ where $h(x_{0})=\pm\frac{\pi}{2},$
$h$ only satisfies a variational inequality.  The condition given
in the definition above
implies that
\begin{equation}\int_{-\infty}^{\infty}[h'v'-\frac{\omega}{2}(1+\nu)\sin{2h}v]
\sech^{m-1}x\,dx\geq0\end{equation}
for any smooth $v$ with compact support which is nonpositive near
$h^{-1}(\frac{\pi}{2})$ and nonnegative near $h^{-1}(-\frac{\pi}{2}).$
This implies that, as a distribution, 
\begin{equation}h''-(m-1)\tanh{x}h'+\frac{\omega}{2}(1+\nu)\sin{2h}=\mu_{+}-\mu_{-},\end{equation}
where $\mu_{+},\mu_{-}$ are positive Radon measures supported on
$h^{-1}(\frac{\pi}{2}),h^{-1}(-\frac{\pi}{2}),$ respectively.
$\mu_{+},\mu_{-}$ are the distributional derivatives of two
monotone functions $F_{+},F_{-}$ which are locally constant outside of
$h^{-1}(\frac{\pi}{2}),h^{-1}(-\frac{\pi}{2}),$ respectively.
This implies that
\begin{equation}h'(x)=C+\int_{0}^{x}[(m-1)\tanh{t}h'-\frac{\omega}{2}(1+\nu)\sin{2h}]dt
+F_{+}-F_{-}.\end{equation}  Thus, $h'$ is continuous, except possibly for
jump discontinuities on $h^{-1}(\frac{\pi}{2}),h^{-1}(-\frac{\pi}{2}),$
with upward jumps on the former and downward jumps on the latter.

If $x_{0}\in h^{-1}(\frac{\pi}{2}),$ then $x<x_{0}$ implies
\begin{equation}\frac{h(x)-h(x_{0})}{x-x_{0}}\geq 0,\end{equation} while $x>x_{0}$
implies 
\begin{equation}\frac{h(x)-h(x_{0})}{x-x_{0}}\leq 0.\end{equation}  The only way this
can be compatible with the fact that $h'$ is only allowed
upward jumps on $h^{-1}(\frac{\pi}{2})$ is if $h'$ is continuous
at $x_{0}$ with
$h'(x_{0})=0.$
A similar argument applies if $x_{0}\in h^{-1}(-\frac{\pi}{2}).$
Hence, $h'$ is continuous on $\real.$

Choose a maximal interval $I$ contained in $h^{-1}((-\frac{\pi}{2},
\frac{\pi}{2})).$  If $I$ is not all of $\real,$ then there is
some endpoint $x_{0}$ contained in either $h^{-1}(\frac{\pi}{2})$
or $h^{-1}(-\frac{\pi}{2}).$  On $I,$ $h$ coincides with a smooth
solution of the Euler-Lagrange equation for $E_{\nu},$ and
extends to $x_{0}$ as a $C^{1}$ function with $h(x_{0})=\pm\frac{\pi}{2}$
and $h'(x_{0})=0.$  But the only solution with this property is
constant, contradicting the assumption that $h$ does not attain
$\pm\frac{\pi}{2}$ as values in $I.$  Hence, $I=\real,$ and
$h$ satisfies the Euler-Lagrange equation everywhere.  \myqed

In order to apply Morse theory, the following result is needed.
\begin{prop}
If $m>1,$ the functional $E_{\nu}:C\rightarrow\real$ satisfies the Palais-Smale
condition, i.e.\ if $(h_{i})\subset C$ is a sequence with
$E_{\nu}(h_{i})$ uniformly bounded and\begin{equation}\lim_{i\rightarrow
\infty} g_{\nu}(h_{i})=0,\end{equation}
then a subsequence of the $h_{i}$
converges strongly to a critical point of $E_{\nu}$ in $C.$  
\end{prop}
\prf  $E_{\nu}$ is a smooth function on $H.$  The fact that
$E_{\nu}(h_{i})$ is uniformly bounded implies that \begin{equation}
\int_{-\infty}^{\infty}(h_{i}')^{2}\sech^{m-1}dx\end{equation}
is uniformly bounded.  Since $h_{i}\in C$ and $m>1,$
this implies that $\parallel h_{i}\parallel_{H}$ is
uniformly bounded so, by passing to a subsequence
if necessary, we can assume that $h_{i}$ converges weakly
in $H$ to some $h\in C.$  By Rellich's Lemma, the restriction
of $(h_{i})$ to any bounded interval $[-a,a]$ is precompact
in $L^{2}([-a,a]).$  Set $h_{0,i}=h_{i}.$
Then, for each positive integer $k,$ we
can choose a subsequence $(h_{k,i})$ of $(h_{k-1,i})$
so that $h_{k,i}$ converges
in $L^{2}([-k,k])$ as $i\rightarrow\infty.$  Replacing $(h_{i})$
by the diagonal subsequence, we may assume that $(h_{i})$ converges
in $L^{2}$ on any bounded interval.  On the other hand, if
$k$ is large enough, then
\begin{equation}\int_{|x|\geq k}|h_{i}|^{2}\sech^{m-1}{x}dx\end{equation}
is as small as we like, since $|h_{i}|\leq\frac{\pi}{2}.$
This implies that\begin{equation}
\int_{-\infty}^{\infty}|h_{i}-h_{j}|^{2}\sech^{m-1}{x}dx\rightarrow 0\end{equation}
as $i,j\rightarrow\infty.$

We can write
\begin{equation}dE_{\nu}(h_{i})(h_{i}-h_{j})-dE_{\nu}(h_{j})(h_{i}-h_{j})
\label{eq:ps}\end{equation}\begin{equation}=
2\int_{-\infty}^{\infty}[(h_{i}'-h_{j}')^{2}-\omega(1+\nu)(\sin{h_{i}}\cos{h_{i}}
-\sin{h_{j}}\cos{h_{j}})(h_{i}-h_{j})]\sech^{m-1}{x}dx.\end{equation}
The second term is bounded in absolute value by
\begin{equation}2\omega\int_{-\infty}^{\infty}(1+\nu)|h_{i}-h_{j}|^{2}\sech^{m-1}{x}dx,\end{equation}
so tends to zero as $i,j\rightarrow\infty.$  
The fact that $g(h_{i})$ tends to zero implies that the
expression in (\ref{eq:ps}) is bounded by arbitrarily small
positive numbers for $i,j\gg 0,$ which implies that 
\begin{equation}\int_{-\infty}^{\infty}(h_{i}'-h_{j}')^{2}\sech^{m-1}{x}dx\end{equation}
tends to zero as $i,j\rightarrow\infty.$  This implies that
the subsequence converges to $h$ in $H.$  \myqed

Recall that a function $F:M\rightarrow\real$ on a Hilbert manifold
is said to be a Morse function if its critical points are isolated
and have nondegenerate Hessians.
Define
\begin{equation}\hat{H}=\{h\in H|h\not\equiv k\pi,h\not\equiv \frac{\pi}{2}+k\pi,
k\in\mathbb{Z}\}.\end{equation}
Theorem 1.1 of \cite{MR93h:58034} implies that, for a generic
set of $\nu$ in the space of compactly supported
$C^{2}$ functions on the line, $E_{\nu}$ is a Morse function on
$\hat{H}.$  In particular, we can choose a sequence $(\nu_{j})$
converging uniformly to zero so that $E_{j}=E_{\nu_{j}}$ is a Morse
function on $\hat{H}$ for each $j.$  When $|\nu|<1,$ it is
straightforward to see that $h\equiv\frac{\pi}{2}+k\pi$ gives
a global minimum for $E_{\nu}$ on $H$ and that the Hessian
is positive definite, so for $j>>0,$ each
$E_{\nu_{j}}$ is a Morse function on the enlarged Hilbert manifold
given by
\begin{equation}\overline{H}=\{h\in H|h\not\equiv k\pi,\,k\in\mathbb{Z}\}.\end{equation}
Note that $h_{\infty}$ is a critical point for any $E_{\nu},$ and
by the result in the previous section, has infinite index iff
it has infinite index as a critical point for $E.$

It will be important for our purposes to use the symmetry
$E_{\nu}(h)=E_{\nu}(-h).$  To exploit this, we will work on
the space $\tilde{C}=(C-\{h_{\infty}\})/\pm.$  $\tilde{C}$ is
a locally convex set in the sense of \cite{MR94e:58023},
Definition 6.2, and each $E_{\nu}$ descends to a smooth function
on $\tilde{C}$ satisfying the Palais-Smale condition on the
subset of $\tilde{C}$ where $E_{\nu}<E_{\nu}(h_{\infty}).$
The basic deformation lemma of Morse theory, adapted to our context,
is the following.
\begin{lemma}
Fix $\nu,$ $\lambda<E_{\nu}(h_{\infty})$ and $\overline{\epsilon}>0.$
Let $C_{\lambda}=E_{\nu}^{-1}((-\infty,\lambda)),$ and
$\tilde{C}_{\lambda}=C_{\lambda}/\pm.$  Let $K_{\lambda}$ be
the set of critical points of $E_{\nu}$ in
$\tilde{C}$ with $E_{\nu}(h)=\lambda,$
and let $N$ be an open neighborhood of $K_{\lambda}$ in $\tilde{C}.$
Then there exist
$\epsilon\in(0,\overline{\epsilon})$ and a continuous map
$\Phi:[0,1]\times\tilde{C}\rightarrow\tilde{C}$ such that
\begin{enumerate}
\item $\Phi(t,h)=h$ if either $t=0,$ $|E_{\nu}(h)-\lambda|\geq
\overline{\epsilon}$ or $h$ is a critical point of $E_{\nu};$
\item $E_{\nu}(\Phi(t,h))$ is nonincreasing as a function of
$t$ for any $h;$ 
\item $\Phi(1,\tilde{C}_{\lambda+\epsilon}-N)\subset
\tilde{C}_{\lambda-\epsilon};$ and
\item $\Phi(1,\tilde{C}_{\lambda+\epsilon})\subset
\tilde{C}_{\lambda-\epsilon}\cup N.$
\end{enumerate}
\end{lemma}
\prf  This is essentially \cite{MR90h:58016}, II,1.9.  The only
difference is that Struwe works there with an actual convex set
rather than the kind of quotient we are dealing with.  Thus,
one has to ensure that the construction of the map $\Phi$
is invariant under $\pm,$ which is straightforward.  
Compare \cite{MR94e:58023}, Theorem 3.3 and \S6.2.  \myqed

The other result from Morse theory we will need is concerned
with the way the topology of $\tilde{C}_{\lambda}$ changes
as $\lambda$ passes a critical value of $E_{\nu}.$  Before
we can apply this, we need to verify the assumption in
\cite{MR90h:58016}, II, 3.3.  Let $h\in\tilde{C}-\{h_{\infty}\}$
be a critical
point for $E_{\nu},$ and let $Q_{h}$ be the Hessian of
$E_{\nu}$ at $h.$  As mentioned previously, $H$ decomposes
into a direct sum 
\begin{equation}H=H_{+}\oplus H_{0}\oplus H_{-},\end{equation}
corresponding to the subspaces on which $Q_{h}$ is positive-definite,
zero and negative-definite, and the latter two subspaces are
finite-dimensional.  In fact, the dimension of $H_{0}$ is at most
1.  That the dimension is at most two follows from the fact that
the relevant differential equation has a 2-dimensional space of solutions
locally; that it is at most 1 follows from the fact that the
two points at infinity fall into the limit point case of Weyl's
classification of singular points for a second order differential
equation.  The assumption we need in order to apply the
theory of \cite{MR90h:58016} is verified by
the following.
\begin{lemma}
If $h\in C-\{h_{\infty}\}$ is a critical point of $E_{\nu},$
then there is an open neighborhood $U$ of $0\in H_{-}$ such that
$h+U\subset C.$  
\end{lemma}
\prf  For any $v,w\in H,$ \begin{equation}Q_{h}(v,w)=
\int_{-\infty}^{\infty}[v'w'-\omega(1+\nu)\cos{2h}\,vw]\sech^{m-1}{x}\,dx
\end{equation}
Fixing $v$ and letting $w$ vary over $H,$ we obtain a bounded
linear functional on $H,$ so there is a bounded linear operator
$A:H\rightarrow H$ such that
\begin{equation}Q_{h}(v,w)=\langle Av,w\rangle_{H}.\end{equation}
$A$ is a symmetric operator, and $H_{-}$ is a direct sum of the
eigenspaces for $A$ corresponding to negative eigenvalues.

Suppose that $Av=\lambda v$ with $\lambda<0.$
Then the fact that $Q_{h}(v,w)=\lambda\langle v,w\rangle_{H}$
for all smooth compactly supported $w$ implies that
\begin{equation}(v'\sech^{m-1}{x})'=(\lambda-1)^{-1}[\lambda+\omega(1+\nu)\cos{2h}]v
\,\sech^{m-1}{x}.\end{equation}  When $x$ is 
sufficiently large in absolute value, $\lambda+\omega(1+\nu)\cos{2h}$
is negative.  This implies that $v'$ is increasing when $v$ is positive,
and is decreasing when $v$ is negative.  Hence, $v$ cannot be zero
when $x$ is large in absolute value.  We can choose a basis
$v_{1},\ldots,v_{N}$ for
$H_{-}$ consisting of eigenfunctions corresponding to eigenvalues
$\lambda_{1},\ldots,\lambda_{N},$ satisfying the condition
that $v_{i}>0$ for $x\gg 0.$ 

Consider $v=\sum_{i}\epsilon_{i}v_{i}$ and suppose that
$\lim_{x\rightarrow\infty}h(x)=\frac{\pi}{2}.$  We need to show that
$h+v\leq\frac{\pi}{2}$ for all sufficiently small $\epsilon_{i}.$
This is true on any bounded interval, so we need only show it is
so when $x\gg0.$  
Let $g=\frac{\pi}{2}-h,$ $w=\sum_{i}v_{i}$
and define the following Wronskian-like
quantity:
\begin{equation}W(x)=[g'(x)w(x) - g(x)w'(x)]\sech^{m-1}{x}.\end{equation}  Then
\begin{equation}\begin{array}{rcl}
W'(x) & = & (g'(x)\sech^{m-1}{x})'w(x)-g(x)(w'(x)\sech^{m-1}{x})'\\
& = & -\frac{\omega}{2}\sin{2h}\,\sech^{m-1}{x}\,w(x)\\ &&-g(x)
\sum_{i}(\lambda_{i}-1)^{-1}[\lambda_{i}+\omega(1+\nu)\cos{2h}]
v_{i}\sech^{m-1}{x}.
\end{array}\end{equation}  This is negative when $x\gg 0.$  On the other hand,
$\lim_{x\rightarrow\infty}W(x)=0,$ so $W(x)>0$ for $x\gg 0.$
This implies that
$(w/g)'<0.$  Hence, $(w/g)<C$ for some $C>0$ and
$x\gg0,$ so the required condition holds when $\epsilon_{i}<C^{-1}$
for each $i.$  A similar argument applies when
$\lim_{x\rightarrow\infty}h(x)=-\frac{\pi}{2}$ or
$x$ tends to $-\infty.$  \myqed

With this in place, we can state the second result from Morse
theory.
\begin{thm}
Suppose $\lambda$ is a critical value of $E_{j}:\tilde{C}
\rightarrow \real$
with
$\lambda<E_{j}(h_{\infty}),$
where, as defined above, $E_j =
E_{\nu_{j}}.$
There are finitely many
critical points $p_{1},\ldots,p_{N}$ in $E_{j}^{-1}(\lambda).$
If the indices of these critical points are are $i_{1},
\ldots,i_{N},$ respectively, then, for any sufficiently
small $\epsilon>0,$ $\tilde{C}_{\lambda+\epsilon}$ is homotopy
equivalent to $C_{\lambda-\epsilon}$ with disks of dimensions
$i_{1},\ldots,i_{N}$ attached along their boundaries.  (If
$i_{k}=0,$ then we add a point to $C_{\lambda-\epsilon}$ as
a disjoint component.)
\end{thm}
\prf  The fact that there are only finitely many critical
points follows from the fact that each critical point
of $E_{j}$ (except possibly $h_{\infty}$) is isolated
together with the Palais-Smale condition.  The rest is
II, Theorem 3.6 in \cite{MR90h:58016}.  \myqed

We are now in a position to prove our main result.  Define
the extended index of a critical point $h$ of $E_{\nu}$ to
be the sum of the dimensions of $H_{-}$ and $H_{0}.$
Since ${\rm dim}\, H_{0} \leq 1,$ the extended index of $h$
is either $i$ or $i+1,$ where $i$
is the index of $h.$
\begin{thm}
Suppose that $F:S^{m}\rightarrow S^{n}$ is an eigenmap with
eigenvalue $\omega,$ $m>1,$ and
\begin{equation}\frac{(m-1)^{2}}{4}<\omega.\end{equation}
There is an infinite sequence $(h_{k})\subset \tilde{C}$ of critical
points for $E$ such that
\begin{enumerate}
\item the extended index of $h_{k}$ is at least $k,$ whereas the
index of $h_{k}$ is at
most k, and
\item the $h_{k}$ converge strongly to $h_{\infty}.$
\end{enumerate}
\end{thm}
\noindent
It should be remembered that each $h_{k}$ corresponds to
a pair of harmonic maps $S^{m+1}\rightarrow S^{n+1}.$
If $F$ is the identity map, then the eigenvalue is $m,$
and the assumption reduces to $2\leq m\leq 5,$ which
gives back the results of \cite{MR97d:58056} and
\cite{MR98d:58039}.\\[.01in]
\prf  $C$ is a contractible space, as is $C-\{h_{\infty}\}.$
This implies that $\tilde{C}$ has the homotopy type of the
classifying space for ${\mathbb{Z}}_{2},$ i.e.\ that of
an infinite-dimensional real projective space.  It follows
that the cohomology ring of $\tilde{C}$ is the polynomial
ring ${\mathbb{Z}}_{2}[x],$ where $x$ has degree 1.  Since
$h_{\infty}$ has infinite index as a critical point of
$E_{j}$ when $j\gg 0,$ we can choose, for any
positive integer $k,$ a $(k+1)$-dimensional subspace $V$
of $H$ on which the Hessian of $E_{j}$ at $h_{\infty}$
is negative definite.  Let $S$ be a small sphere in $V$
centered at the origin;
when $S$ is sufficiently small, $E_{j}$ takes values strictly
less than $E_{j}(h_{\infty})$ on $h_{\infty}+S.$  This implies
that the nontrivial homology class in $\tilde{C}$ of degree
$k$ exists already in some $\tilde{C}_{\lambda}$ with
$\lambda<E_{j}(h_{\infty}).$  As described in the previous
Theorem, the homotopy type of $\tilde{C}_{\lambda}$ is obtained
by taking a point (corresponding to the unique global minimum of
$E_{j}$ in $\tilde{C})$ and attaching disks of dimensions
determined by the indices of critical points with energies
less than $\lambda.$  The only way to create a homology
class of degree $k$ is by attaching a disk of dimension $k.$
Hence, $E_{j}$ must have at least one critical point of every
possible index, for each $j\gg 0.$

Now choose a critical point $f_{j}$ of $E_{j}$ of index $k$ for
each $j\gg 0.$  We will show that $(f_{j})$ satisfies the hypothesis
of the Palais-Smale condition for $E.$  We know that
\begin{equation}E_{j}(f_{j})<E_{j}(h_{\infty})=\frac{\omega}{2}
\int_{-\infty}^{\infty}(1+\nu_{j})\sech^{m-1}{x}\,dx
\leq\omega \int_{-\infty}^{\infty}\sech^{m-1}{x}\,dx,\end{equation}
so is bounded independent of $j.$
On the other hand, the fact that $\nu_{j}$ converges uniformly
to zero implies that
\begin{equation}E_{j}(f_{j})=\frac{1}{2}\int_{-\infty}^{\infty}
[(f_{j}')^{2}+\omega(1+\nu_{j})\cos^{2}{f_{j}}]\sech^{m-1}{x}\,dx\end{equation}
\begin{equation}\geq\frac{1}{4}\int_{-\infty}^{\infty}[(f_{j}')^{2}+
\omega\cos^{2}{f_{j}}]\sech^{m-1}{x}\,dx=\frac{1}{2}E(f_{j})\end{equation}
when $j\gg 0.$
Thus, $E$ is uniformly bounded on the sequence $(f_{j}).$
On the other hand, $f_{j}$ satisfies the Euler-Lagrange
equation
\begin{equation}f_{j}''-(m-1)\tanh{x}f_{j}'+\frac{\omega}{2}(1+\nu_{j})
\sin{2f_{j}}=0,\end{equation}
which implies that
\begin{equation}dE(f_{j})(v)=-2\int_{-\infty}^{\infty}[f_{j}''-(m-1)
\tanh{x}f_{j}'+\frac{\omega}{2}\sin{2f_{j}}]v\sech^{m-1}{x}\,dx\end{equation}
\begin{equation}=-\omega\int_{-\infty}^{\infty}\nu_{j}v\sin{2f_{j}}\,\sech^{m-1}{x}\,dx.\end{equation}
The integral on the last line is bounded in absolute value by
\begin{equation}\omega\parallel\nu_{j}\parallel_{C^{0}}\int_{-\infty}^{\infty}|v|
\sech^{m-1}{x}\,dx.\end{equation}  This tends to zero as $j\rightarrow\infty,$
so $(f_{j})$ satisfies the Palais-Smale condition.  We can thus
choose a subsequence which converges in $H$ to some some
critical point $h_{k}$ of $E$ in $C.$

We need to show that $h_{k}\not=h_{\infty},$ and that its
extended index is at least $k.$  Define $c_{k}$ to be
the infimum of all $\lambda$ such that the nontrivial homology
class of degree $k$ in $\tilde{C}$ can be represented by
a cycle in $E^{-1}((-\infty,\lambda)).$  From the fact that
the class can be represented as described above by an embedding
of a real projective space of dimension $k$ in $\tilde{C}$ along
which the energy is everywhere less than $E(h_{\infty}),$ it
follows that $c_{k}<E(h_{\infty}).$
The fact that
\begin{equation}|E_{j}(h)-E(h)|=|\frac{1}{2}\int_{-\infty}^{\infty}
\nu_{j}\cos^{2}{h}\,\sech^{m-1}{x}\,dx|\leq C\sup_{x\in\real}|\nu_{j}
(x)|\end{equation}
implies that $E_{j}$ converges uniformly to $E$ on $H.$
Thus, for any $\epsilon\in(0,\frac{1}{4}(
E(h_{\infty})-c_{k})),$ $j\gg 0$ implies
that
\begin{equation}E^{-1}((-\infty,c_{k}+\frac{\epsilon}{2}))\subset
E_{j}^{-1}((-\infty,c_{k}+\epsilon))\end{equation}
and $c_{k}+\epsilon<E_{j}(h_{\infty}).$  But this means
that $E_{j}^{-1}((-\infty,c_{k}+\epsilon))$ must contain
some critical point of index $k.$  We can therefore assume
that $E_{j}(f_{j})<c_{k}+\epsilon,$ which would imply that
$E(h_{k})<c_{k}+\epsilon.$  This shows that $h_{k}\not=h_{\infty}.$

To see that the extended index of $h_{k}$ is at least
$k,$ we can look at the difference between the Hessians of
$E_{j}$ and $E$ at $f_{j}$ and $h_{k},$ respectively.
We find
\begin{equation}D^{2}E_{j}(f_{j})(v,w)-D^{2}E(h_{k})(v,w)=
\frac{\omega}{2}\int_{\infty}^{\infty}[\cos{2h_{k}}-
(1+\nu_{j})\cos{2f_{j}}]vw\sech^{m-1}{x}\,dx.\end{equation}
This tends to zero as $j\rightarrow\infty,$ uniformly
in $v,w$ as they range over any bounded set in
$H.$  This implies that the Hessian of $E_{j}$ at $f_{j}$
converges to that of $E$ at $h_{k}.$  The extended index
is upper semicontinuous on the space of continuous quadratic forms
on $H,$ so the extended index of $h_{k}$ is at least $k.$ 
Similarly, the index is lower semicontinuous, so the index cannot be
greater than k.

Finally, the sequence of $h_{k}$ satisfies the hypothesis of
the Palais-Smale condition, so converges to some critical point
of $E.$  By an argument similar to the one just given, the
limit must have infinite index, so the limiting critical
point must be $h_{\infty}.$  \myqed

The analogous argument can be carried out for $C^{+}$ and
$C^{-}.$  This leads to the following conclusion.
\begin{thm}
Suppose that $F:S^{m}\rightarrow S^{n}$ is an eigenmap with
eigenvalue $\omega,$ $m>1,$ and
\begin{equation}\frac{(m-1)^{2}}{4}<\omega.\end{equation}
There are infinite sequences of critical
points for $E$ in $C^{+}$ and $C^{-},$
each of which converges strongly to $h_{\infty}.$
\end{thm}
These are the generalizations of the infinite sequences of
degree 0 and degree 1 harmonic maps found in \cite{MR97d:58056}
and \cite{MR98d:58039}.  It is of interest to ask which homotopy
classes of maps between spheres can be represented as suspensions
of eigenmaps of spheres.  As mentioned previously, the
Hopf maps $S^{3}\rightarrow S^{2}$ and $S^{7}\rightarrow S^{4}$
are eigenmaps and satisfy
the hypothesis of Theorem 2.1.  Therefore, the homotopy classes
of their suspensions contain infinitely many harmonic representatives.
In the case of the map $S^{3}\rightarrow S^{2},$ we obtain a
map representing the nontrivial class in $\pi_{4}(S^{3})={\mathbb Z}_{2}.$
In the case of $S^{7}\rightarrow S^{4},$ we obtain a generator
of $\pi_{8}(S^{5})={\mathbb Z}_{24}.$
The maps $S^{5}\rightarrow S^{4}$ and $S^{9}\rightarrow S^{8}$
mentioned in \S2 produce infinite families of harmonic maps
of the form $S^{6}\rightarrow S^{5}$ and $S^{10}\rightarrow S^{9}.$
The relevant homotopy groups are again isomorphic to ${\mathbb Z}_{2},$
but we do not know whether the suspensions of the two original
maps represent the nontrivial class.  The maps associated to
the cubic harmonic eiconals on ${\mathbb R}^{8}$ and ${\mathbb R}^{14}$
give infinite sequences of harmonic self-maps of degrees 0 and 2 
defined on $S^{8}$ and $S^{14}.$

As we have already
mentioned, there are other settings where similar ideas may apply.  We will
briefly summarize the characteristics of the problem discussed
here which make the argument above possible.
\begin{enumerate}
\item The configuration space is contractible, being in this
case a Hilbert space.
\item The energy functional satisfies the Palais-Smale condition.
\item There is a reflection symmetry of the configuration space preserving
the energy functional.
There is a unique fixed point for this symmetry,
corresponding to a critical point for the energy.
\item The index of the fixed point is infinite.
\item All critical points with energy less than
that of the fixed point have
finite index.
\item 
Possibly after small perturbations of the energy, the critical points
with energy less than that of the fixed point are nondegenerate.
\end{enumerate}
Of course, there are variations of these conditions which
may be treated along similar lines.

\end{document}